\documentclass[letterpaper, aps, tightenlines, nofootinbib, 11pt]{article}
\pdfoutput = 1

\usepackage{setspace}
\usepackage[bookmarks = false, colorlinks = true, linkcolor = blue, citecolor = purple]{hyperref}
\usepackage{shortcuts}
\usepackage{titlesec}
\usepackage{cite}
\usepackage{tocloft}

\usepackage{tikz}
\usetikzlibrary{arrows,decorations.pathmorphing,backgrounds,positioning,fit,petri}

\usepackage[margin = 2.5cm]{geometry}
\titleformat{\section}
  {\normalfont\large\bfseries}{\thesection}{1em}{}
\titleformat{\subsection}
  {\normalfont\large\it}{\thesubsection}{1em}{}
\titleformat{\subsubsection}
  {\normalfont\it}{\thesubsubsection}{1em}{}
\numberwithin{equation}{section}

\begin{document}

\begin{center}

\thispagestyle{empty}

\begin{flushright}
\texttt{SU-ITP-14/08}
\end{flushright}

\vspace*{5em}

{ \LARGE \bf Notes on Entanglement in Abelian Gauge Theories}

\vspace{1cm}

{\large \DJ or\dj e Radi\v cevi\'c}
\vspace{1em}

{\it Stanford Institute for Theoretical Physics and Department of Physics\\ Stanford University \\
Stanford, CA 94305-4060, USA}\\
\vspace{1em}
\texttt{djordje@stanford.edu}\\

\vspace{0.08\textheight}
\begin{abstract}
We streamline and generalize the recent progress in understanding entanglement between spatial regions in Abelian gauge theories. We provide an unambiguous and explicit prescription for calculating entanglement entropy in a $\Z_N$ lattice gauge theory. The main idea is that the lattice should be split into two disjoint regions of links separated by a buffer zone of plaquettes. We show that the previous calculations of the entanglement entropy can be realized as special cases of our setup, and we argue that the ambiguities reported in the previous work can be understood as basis choices for gauge-invariant operators living in the buffer zone. The proposed procedure applies to Abelian theories with matter and with continuous symmetry groups, both on the lattice and in the continuum.
\end{abstract}
\end{center}

\newpage

{\small
\tableofcontents
}

\section{Introduction and summary}

Entanglement between spatial regions in gauge theories is a subtle subject because the Hilbert space of gauge-invariant states cannot be factored into a direct product of Hilbert spaces defined on each region. In general, reduced density matrices and entanglement entropies in lattice gauge theories can be computed by relaxing the gauge constraint, or equivalently by minimally embedding the physical Hilbert space in a larger space which does factorize into a direct product of states on each region \cite{Levin:2006zz, Buividovich:2008yv, Donnelly:2011hn, Donnelly:2012st, Buividovich:2008gq}. These methods yield sensible results, but they depend on assuming that the minimal extension of the Hilbert space is the correct one; it is natural to ask whether one can calculate gauge theory entanglement entropies without extending the Hilbert space.  This is indeed possible in certain special cases, both on the lattice and in the continuum \cite{Hamma:2005zz, Kitaev:2005dm, Agon:2013iva, Gromov:2014kia, Velytsky:2008rs, Yao:2010} (also see \cite{Eisert:2008ur} for a review of some condensed-matter models for which such computations are available), and for certain conformal theories in particular a vast amount of holographic results can be found by using the Ryu-Takayanagi prescription \cite{Ryu:2006bv} (for a review, see \cite{Nishioka:2009un}). What can be said about a general setup, though?

An insightful recent paper by Casini, Huerta, and Rosabal has highlighted that, in general, the need for auxiliary degrees of freedom disappears when the entanglement entropy in a lattice gauge theory is defined using a subalgebra of operators defined on one subset of lattice links \cite{Casini:2013rba}.\footnote{Lattice bipartitions in which the entanglement boundary did not cut through links have appeared before, e.g.~in \cite{Hamma:2005zz}.} The argument goes as follows. Let $V$ be a subset of links of a given lattice, and let $\A_V$ be an algebra of gauge-invariant operators supported on links in $V$ (this need not be the maximal possible such algebra). The Gauss law implies that $\A_V$ has a nontrivial center $\mathcal Z_V$, which is also the center of the algebra of gauge-invariant operators constructed on links $\bar V$ that are not in $V$. In the basis in which all $\mathcal Z_V$ generators are diagonal, the physical Hilbert space can be split into superselection sectors,
\bel{
  \H = \bigoplus_{\b k} \left(\H^{(\b k)}_V \otimes \H^{(\b k)}_{\bar V} \right),
}
where $\b k$ is the set of eigenvalues of center generators. By tracing over the external degrees of freedom one can then construct a reduced density matrix $\rho_V$ as a direct sum of reduced density matrices in each sector. This matrix acts on the set of states $\H_V = \bigoplus_{\b k} \H^{(\b k)}_V$ that all live just in $V$. This way one explicitly constructs the operator $\rho_V$ for which $\Tr_{\H_V} \left(\rho_V \O\right)$ computes $\avg \O$, the expectation of \emph{any} $\O \in \A_V$ in the full theory on the entire lattice. The associated entanglement entropy is then computed as the von Neumann entropy of $\rho_V$. All the older prescriptions for calculating the entanglement entropy via embedding are specific special cases of this procedure.

Ref.~\cite{Casini:2013rba} left unresolved a few ambiguities; in particular, their definitions of different choices of algebras $\A_V$ were found to lead to different values of even the universal piece in the entanglement entropy. In this note we resolve these ambiguities for Abelian gauge theories. We show that all algebras $\A_V^\alpha$ appropriate for a subset $V$ of the lattice lead to the same results for the entanglement entropy. Furthermore, we argue that the entanglement entropy in a lattice gauge theory is naturally computed by splitting the lattice into two disjoint sets of links separated by a region (\emph{buffer zone}) exactly one lattice spacing wide. Different choices of algebras (also called superselection rules, boundary conditions, or centers) are simply found to correspond to different choices of bases for the degrees of freedom living in the buffer zone, and this is why the entanglement entropy is independent of the algebra choice.

This paper is organized such that the build-up to our conclusion is gradual and stems from the framework established in \cite{Casini:2013rba}. In Sec.~\ref{sec setup} we introduce our notation and set up some elementary $\Z_N$ lattice gauge theory framework. In Sec.~\ref{sec ebc} we compute the reduced density matrix on a region $V$ with a particular choice of the algebra, $\A_V^e$, that is associated to imposing electric boundary conditions; this construction is related to that of the ``electric center'' presented in \cite{Casini:2013rba}. In  Sec.~\ref{sec mbc} we introduce a class of related algebras $\A_V^\alpha$ and focus on a particular one, $\A_V^m$, which is associated to magnetic boundary conditions (related but inequivalent to the ``magnetic center'' of \cite{Casini:2013rba}). Our punchline comes in Sec.~\ref{sec embc}. We first show that the set $\{\A_V^\alpha\}$ contains all algebras that can be associated to the lattice subset $V$, and next we show that all the boundary conditions defined in Secs.~\ref{sec ebc} and \ref{sec mbc} can be viewed as different choices of basis for the same set of degrees of freedom living in the buffer zone at the boundary of $V$. Therefore, the entanglement entropies $S_V^\alpha$ associated to the algebras $\A_V^\alpha$ are all equal, and in the case of a topological state they correctly reproduce the expected results.  In Secs.~\ref{sec u1}, \ref{sec matter}, and \ref{sec continuum} we show that our proposal has natural generalizations to theories with a continuous gauge group, theories with matter, and theories in a continuum. We finish with an outline of open research directions.

\section{$\Z_N$ lattice gauge theory}\label{sec setup}

Consider a Hamiltonian formulation of a $\Z_N$ lattice gauge theory in $d$ spatial dimensions. We work on a finite lattice with no non-trivial topology and with open boundary conditions. Let $g$ be the generator of the $\Z_N$ transformation, so that $g^N = 1$. The $N_S$ lattice sites are labeled by $i, j, \ldots$, and the $N_L$ links are labeled either by a link index, $\ell$, a pair of adjacent site indices, $(i,j)$, or by a site and a direction, $(i,\mu)$. The quantum variables $U$ live on links and take values in $\Z_N$, so a single link Hilbert space is isomorphic to $\C^N$. The operator algebra on this space is generated by the \textbf{momentum operator} $ L_\ell$ and the \textbf{position operator} $ U^r_\ell$, which act on $\qvec U_\ell$ via
\bel{\label{pos mom ops}
   L_\ell \qvec U_\ell = \qvec {g U}_\ell,\quad  U^r_\ell \qvec U_\ell = r(U) \qvec U_\ell,
}
where $r$ is a representation of the gauge group. Lattice links are directed, so if $\ell = (i,j)$, the link of opposite direction is ${\ell^{\trm T}} = (j, i)$, and it is convenient to define $ L_{{\ell^{\trm T}}}$ and $ U^r_{{\ell^{\trm T}}}$ via
\bel{
   L_{{\ell^{\trm T}}} \qvec U_\ell = \qvec{U g^{-1}}_\ell, \quad  U^r_{{\ell^{\trm T}}} \qvec U_\ell = r(U^{-1}) \qvec U_\ell.
  }
Finally, a state of the entire lattice system is denoted by $\qvec U$, and it belongs to a Hilbert space $\H_0 \simeq \bigotimes_\ell (\C^N)_\ell$, with operators that are elements of the algebra $\A_0 \simeq \bigotimes_\ell GL(\C, N)_\ell$. Operators on different links all commute. Henceforth, whenever we write $L_\ell$ (for instance), we refer to the operator on the full lattice $L_\ell \otimes \left(\bigotimes_{\ell' \neq \ell} \1_{\ell'}\right)$.

We are interested in systems invariant under the local transformation
\bel{
  U_{ij} \mapsto U_{ij}^\Lambda \equiv g^{\Lambda_i} U_{ij} g^{- \Lambda_j}.
}
This transformation is implemented in terms of operators via
\bel{
  \qvec{U^\Lambda} \equiv \prod_i  G_i^{\Lambda_i} \qvec U \equiv \prod_{i,\, \mu} \left( L_{i\mu}\right)^{\Lambda_i} \qvec U.
}
In the above product there are exactly two momentum operators acting on each link, one acting in the direction of the link and one in the opposite direction. Together they implement the desired local transformation. Gauging this transformation, we demand that physical states are only those satisfying $\qvec{U^\Lambda} = \qvec U$. This gauge-invariance constraint can be written as the Gauss law,
\bel{\label{gauss}
   G_i\qvec U = \prod_{\mu}  L_{i\mu}\qvec U = \qvec U.
}
Gauge-invariant operators (those that preserve gauge-invariance of states) are all momentum operators and all loops of position operators, so the gauge-invariant algebra $\A \subset \A_0$ is generated by operators that act as
\bel{\label{em gens}
   L_\ell \qvec U_\ell = \qvec {g U}_\ell,\quad  W^r_p \qvec {U} = \prod_{\ell \in p} r(U_\ell) \qvec{U},
}
where $p$ is any plaquette. We will refer to them as \textbf{electric} and \textbf{magnetic generators}, respectively.

The Gauss law \eqref{gauss} takes on a particularly simple form in the electric basis, where $ L_\ell$ is diagonalized on each link. The eigenstates on one link are
\bel{\label{el basis}
  \qvec k_\ell \equiv \frac1{\sqrt N}\sum_{n = 0}^{N - 1} e^{-2\pi i n k /N} \qvec{g^n}_\ell, \qquad  L_\ell \qvec k_\ell =  e^{2\pi i k/N} \qvec k_\ell.
}
Here we define \textbf{electric fluxes} $k = 0, \ldots, N-1$. Under a gauge transformation, these states transform as $\qvec k_\ell \mapsto e^{2\pi i k \del_\ell \Lambda/N } \qvec k_\ell$, where $\del_\ell$ is the difference operator between sites connected by $\ell$. The Gauss law now states that a gauge-invariant state has a net zero flux flowing into each lattice site. Evidently, the Hilbert space of physical states, $\H \subset \H_0$, is spanned by loops of constant electric flux, created and annihilated by the magnetic generators $ W^r_p$. As with the links, each individual plaquette admits a Hilbert space isomorphic to $\C^N$, and so we can view the physical Hilbert space as a direct product over all plaquettes,
\bel{\label{H}
  \H \simeq \bigotimes_p (\C^N)_p.
}
From here on we will always work with the representation $r(g) = e^{2\pi i/N}$.

\section{Entanglement with electric boundary conditions} \label{sec ebc}

Consider partitioning the set of links into two disjoint subsets, $V$ and $\bar V$, such that each link is either in $V$ or in $\bar V$. The set $\H$ of physical states over $V \cup \bar V$ cannot factorize into a tensor product of physical Hilbert spaces over $V$ and over $\bar V$ due to the Gauss law constraint \eqref{gauss}. To deal with this, we follow the spirit, if not the letter, of \cite{Casini:2013rba}. Let us focus on gauge-invariant operator subalgebras (\textbf{electric algebras}) $\A_V^e$ and $\A_{\bar V}^e$ generated by electric and magnetic generators \eqref{em gens} defined on all links and plaquettes that wholly lie in $V$ and $\bar V$, respectively. Consider a lattice site $i$ with links from both $V$ and $\bar V$ emanating from it; the set of all such sites is the \textbf{entangling boundary} $\del V$. We will often refer to ``links in $\del V$;'' by this we refer to the chains of $V$ links connecting adjacent sites in $\del V$.
 
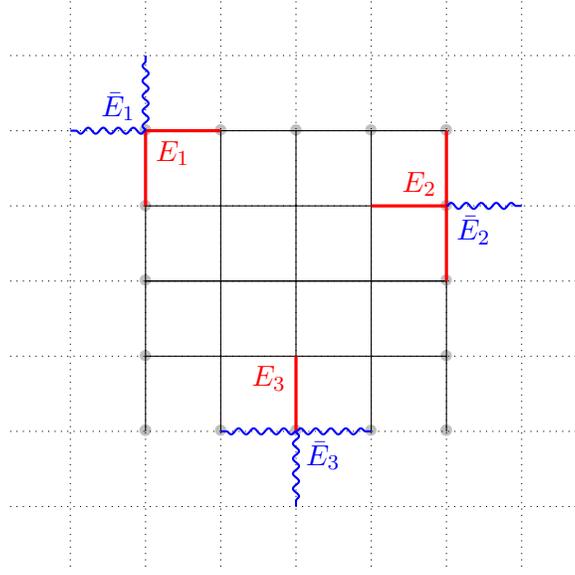
\begin{figure}
\begin{center}

\begin{tikzpicture}[scale = 2]
  \foreach \x in {-1, -0.5,..., 1} \draw (\x, -1) node[lightgray] {$\bullet$};
  \foreach \x in {-1, -0.5,..., 1} \draw (\x, 1) node[lightgray] {$\bullet$};
  \foreach \x in {-1, -0.5,..., 1} \draw (1, \x) node[lightgray] {$\bullet$};
  \foreach \x in {-1, -0.5,..., 1} \draw (-1, \x) node[lightgray] {$\bullet$};

  \draw[step = 0.5, dotted] (-1.9, -1.9) grid (1.9, 1.9);
  \draw[step = 0.5] (-1, -0.5) grid (1, 1);
  \draw[xstep = 0.5, ystep = 5] (-1, -1) grid (1, -0.5);

  \draw[very thick, red] (-1, 0.5) -- (-1, 1) -- (-0.5, 1);
  \draw[red] (-1, 1) node[anchor = north west] {$E_1$};
  \draw[thick, blue, decorate, decoration = {snake, amplitude = .4mm, segment length = 2mm, post length = 0.1mm}] (-1, 1) -- (-1.5, 1);
  \draw[thick, blue, decorate, decoration = {snake, amplitude = .4mm, segment length = 2mm, post length = 0.1mm}] (-1, 1) -- (-1, 1.5);
  \draw[blue] (-1, 1) node[anchor = south east] {$\bar E_1$};

  \draw[very thick, red] (1, 0.5) -- (0.5, 0.5);
  \draw[very thick, red] (1, 0.5) -- (1, 1);
  \draw[very thick, red] (1, 0.5) -- (1, 0);
  \draw[red] (1, 0.5) node[anchor = south east] {$E_2$};
  \draw[thick, blue, decorate, decoration = {snake, amplitude = .4mm, segment length = 2mm, post length = 0.1mm}] (1, 0.5) -- (1.5, 0.5);
  \draw[blue] (1, 0.5) node[anchor = north west] {$\bar E_2$};

  \draw[very thick, red] (0, -0.5) -- (0, -1);
  \draw[red] (0, -0.5) node[anchor = north east] {$E_3$};
  \draw[thick, blue, decorate, decoration = {snake, amplitude = .4mm, segment length = 2mm, post length = 0.1mm}] (0, -1) -- (0, -1.5);
  \draw[thick, blue, decorate, decoration = {snake, amplitude = .4mm, segment length = 2mm, post length = 0.1mm}] (0, -1) -- (0.5, -1);
  \draw[thick, blue, decorate, decoration = {snake, amplitude = .4mm, segment length = 2mm, post length = 0.1mm}] (0, -1) -- (-0.5, -1);
  \draw (0, -1) node[blue, anchor = north west] {$\bar E_3$};
\end{tikzpicture}

\end{center}
\caption{\small (\textsc{color online}) Three examples of boundary electric operators defined in eq.~\eqref{bdry el ops}. Full lines denote links in $V$ and dotted lines denote links in $\bar V$. Gray dots denote elements of $\del V$. The boundary electric operators $E_i$, defined on each boundary site, are products of electric generators on all links that enter that site and belong to $V$ (thick red lines). Similarly, boundary electric operators $\bar E_i$ are products of electric generators on all links entering site $i$ and belonging to $\bar V$ (wavy blue lines). The pictured choice of $(V, \bar V)$ has 16 boundary lattice sites (i.e.~elements of $\del V$) and equally many pairs $(E_i, \bar E_i)$.}
\label{fig el bdry ops}
\end{figure}

Due to the Gauss law, $G_i = \prod_\mu  L_{i\mu} = 1$ holds for all physical states. Each electric generator that enters a given $G_i$ belongs to either $\A_V^e$ or $\A_{\bar V}^e$. Hence we can write the Gauss law for site $i$ in the generic form
\bel{
   E_i =  \bar E_i^{-1},
}
with
\bel{\label{bdry el ops}
    E_i \equiv \prod_{(i,\, j) \in V}  L_{ij}, \quad \bar E_i \equiv \prod_{(i,\, j) \in \bar V}  L_{ij}
}
being the \textbf{boundary electric operators}; their eigenvalues measure the total electric flux that flows out of site $i$ from either side of the partition. See Fig.~\ref{fig el bdry ops} for several examples.

Thus, for each boundary site, the boundary electric operator $E_i \in \A_V^e$ can be expressed as an operator $\bar E_i^{-1} \in \A_{\bar V}^e$, and vice versa. Since all elements of $\A_{\bar V}^e$ commute with all elements of $\A_V^e$, all boundary  electric operators belong to the joint center of $\A_V^e$ and $\A_{\bar V}^e$. Consequently, we will say that the algebra generated by all the electric boundary operators is the \textbf{electric center} $\mathcal Z^e_V$ of both $\A_V^e$ and $\A_{\bar V}^e$. There are $|\del V|$ electric boundary generators, but typically not all of them are independent; we will discuss constraints in a moment.

Elements of the electric center will not commute with all magnetic generators on plaquettes that contain links in both $V$ and $\bar V$. It is for the lack of such ``boundary magnetic operators'' that the algebra $\A^e$ generated by $\A_V^e \cup \A_{\bar V}^e$ differs from the full algebra $\A$. By altering the above notion of bipartition, it is possible to construct algebras $\A_{V}^\alpha$ and $\A_{\bar V}^\alpha$ whose mutual center is also generated by a number of magnetic generators. Indeed, we will show in the next section that a purely magnetic center can be constructed. In this section we focus on the simplest case of the pure electric center.

All boundary electric operators \eqref{bdry el ops} are diagonal in the electric basis \eqref{el basis}. For a given state in this basis, the eigenvalues of the $E_i$ operators are  $e^{2\pi i k_i/N}$, where $k_i$ for $i \in \del V$ measures the electric flux flowing out of $\bar V$ into $V$. Correspondingly, the eigenvalues of the $\bar E_i$'s are $e^{-2\pi i k_i/N}$. The multiplet of fluxes $\b k \equiv (k_1, \ldots, k_{|\del V|})$ flowing out of all boundary vertices thus labels $N^{|\del V|- n_\del}$ superselection sectors, where $n_\del$ is the number of connected components of $V$ (there is one constraint per such component, as the net flux flowing out of any region must be zero).

The existence of superselection sectors means that $\H$ (i.e.~the set of electric eigenstates with flux running along loops, as discussed around eq.~\eqref{H}) can be written as a direct sum of $\H^{(\b k)}$'s, the spaces of physical states with boundary electric operators having eigenvalues $\b k$. This is just the sum over electric boundary conditions (electric field configurations) on $\del V$. Now each $\H^{(\b k)}$ is individually decomposed into the tensor product of states on $V$ and $\bar V$ with the given incoming (or outgoing for the case of $\bar V$) flux $\b k$, so we have the decomposition
\bel{
  \H = \bigoplus_{\b k }\H^{(\b k)} = \bigoplus_{\b k}\left( \H^{(\b k)}_V \otimes \H^{(\b k)}_{\bar V} \right).
}
We will henceforth let $D^{(\b k)} \equiv \dim \H^{(\b k)}$, and correspondingly $D^{(\b k)}_V = \dim \H^{(\b k)}_V$ and $D^{(\b k)}_{\bar V} = \dim \H^{(\b k)}_{\bar V}$. Note that, according to \eqref{H}, $D \equiv \dim \H = N^{N_P}$ where $N_P$ is the number of plaquettes. The space of states on $V$ is thus defined as $\H_V = \bigoplus_{\b k} \H_V^{(\b k)}$, and here we see the formal justification of the intuition that ``the Gauss law should be relaxed'' when talking about entanglement \cite{Buividovich:2008gq, Agon:2013iva, Donnelly:2011hn}.

Any operator in $\A^e$ can be written in block-diagonal form with $N^{|\del V|-n_\del}$ blocks, each block being a square matrix with $D^{(\b k)} = D^{(\b k)}_V  D^{(\b k)}_{\bar V}$ rows.  The difference between $\A$ and $\A^e$ is generated by Wilson loops (magnetic generators) around plaquettes containing links in both $V$ and $\bar V$. These operators mix superselection sectors in a simple way: a magnetic generator that connects two vertices $i, j \in \del V$ shifts the superselection sector from the one with $(k_i, k_j)$ to one with $(k_i \pm 1, k_j \mp 1)$, depending on the direction of the loop.

If the entire lattice is in a  state described by a density matrix $\rho$, the expectation value of an operator $\O \in \A_V^e$ is
\bel{
  \avg{\O} = \Tr_{\H} \left(\rho \O\right) = \sum_{\b k} \Tr_{\H^{(\b k)}} \left(\rho \O\right).
}
As $\O$ maps $\H^{(\b k)}$ to itself, the only elements of the density matrix that enter the sum are those that belong to the same block-diagonal form found above. Thus, for computing the reduced density matrix, we may replace $\rho$ by its block-diagonal restriction $\~\rho$  defined by
\begin{equation}
  \qmat {U_1}{\~\rho}{U_2} \equiv
    \Bigg\{
      \begin{array}{ll}
      \qmat {U_1}{\rho}{U_2}, & \hbox{$U_1$ and $U_2$ are in the same sector;} \\
      0, & \hbox{else.}
      \end{array}
\end{equation}
In addition, we define
\algnl{
  \~\rho &\equiv \bigoplus_{\b k}\ p_{\b k}\ \rho^{(\b k)},\\ \label{def pk}
  p_{\b k} &\equiv \Tr_{\H^{(\b k)}} \rho.
}
With this definition all the matrices $\rho^{(\b k)}$ have unit trace and can be interpreted as density matrices themselves. The $p_{\b k}$ measure the trace of each block in $\rho$, and they obey $\sum_{\b k} p_{\b k} = 1$.

In each superselection sector the Hilbert space factorizes, so we can trace out the $\bar V$ degrees of freedom in the usual way, getting the density matrix appropriate to electric boundary conditions,
\bel{
  \avg{\O} = \sum_{\b k} p_{\b k} \Tr_{\H_V^{(\b k)}} \left(\rho_V^{(\b k)} \O\right) \equiv \Tr_{\H_V} \left(\rho_V^e \O\right),
}
with
\algnl{
  \rho^{(\b k)}_V &\equiv \Tr_{\H^{(\b k)}_{\bar V}} \rho^{(\b k)},\\
  \rho_V^e &\equiv \bigoplus_{\b k}\  p_{\b k}\  \rho^{(\b k)}_V,\\ \label{HV}
  \H_V &\equiv \bigoplus_{\b k} \H^{(\b k)}_V.
}
We have now explicitly constructed the matrix $\rho_V^e$ that can be used to compute the expectation value of any $\O \in \A_V^e$ by tracing out just the degrees of freedom in $\H_V$. This is precisely the reduced density matrix of interest, and the entanglement entropy can now be found as the von Neumann entropy of $\rho_V^e$, leading to the central equation in these notes:
\algnl{\label{SV}
  S_V^e
  &= - \Tr_{\H_V}\left(\rho_V^e \log \rho_V^e\right) = H(\{p_{\b k}\}) + \sum_{\b k} p_{\b k} S_V^{(\b k)},
}
with
\algnl{
  H(\{p_{\b k}\}) &\equiv -\sum_{\b k} p_{\b k}\log p_{\b k},\\
  S_V^{(\b k)} &\equiv  \Tr_{\H^{(\b k)}_V} \left(\rho_V^{(\b k)} \log \rho_V^{(\b k)} \right).
}
This equation has already been featured in \cite{Casini:2013rba}, and in a slightly different form in \cite{Donnelly:2011hn}.

As a simple example, if we start in a pure state $\qvec {U_0}$ in a definite electric sector $\b k_0$, the initial density matrix will be $\rho = \qvec {U_0} \qvecconj{U_0}$. From eq.~\eqref{def pk} we have $p_{\b k} = \delta_{\b k, \b k_0}$, and eq.~\eqref{SV} readily reveals that the associated entanglement entropy is
\bel{
  S_V^e[U_0] = S_V^{(\b k_0)} = - \Tr_{\H^{(\b k_0)}_V}\left(\rho_V^{(\b k_0)} \log \rho_V^{(\b k_0)}\right).
}
Thus, for any state for which the boundary electric fluxes are good quantum numbers, the entanglement entropy is evaluated as in any other field theory, with the single caveat that the fields being traced out must all be restricted to the correct boundary condition at $\del V$ provided by the starting state.

As another example, consider the pure topological state $\qvec{\trm{topo}}$ which is a superposition of all gauge-invariant electric basis vectors, i.e.~for which $\rho = \frac1{D} \b 1_D$, where $D = N^{N_P}$ is the dimension of $\H$ and $\b 1_D$ is a $D \times D$ matrix whose each entry is one (not to be confused with the identity matrix $\1_D$).\footnote{The topological nature of this state is (heuristically) easy to see: if we gauge the loop group, i.e.~the ``diffeomorphisms'' that take a state specified by some electric flux loop to a state specified by a different flux loop, the only gauge-invariant state will be the topological state $\qvec{\trm{topo}}$.  This is the ground state of the famous toric code \cite{Kitaev:2005dm}, and in the language of \cite{Levin:2006zz} it can also be viewed as a condensate of electric strings.} We can now once more turn the crank \eqref{def pk} and find $\rho^{(\b k)} = \frac1{D^{(\b k)}} \b 1_{D^{(\b k)}}$ and $p_{\b k} = \frac{D^{(\b k)}}{D}$. Tracing over $\bar V$ links gives $\rho^{(\b k)}_V = \frac1{D_V^{(\b k)}} \b 1_{D_V^{(\b k)}}$. This is a density matrix for a pure state, so $S_V^{(\b k)} = 0$ and
\bel{\label{SV topo}
  S_V^e[\trm{topo}] = - \sum_{\b k} p_{\b k} \log p_{\b k} = \log D - \sum_{\b k} \frac{D^{(\b k)}}{D} \log D^{(\b k)}.
}
In case all superselection sectors have the same number of elements (this is always true for sufficiently simple lattices), we have $D^{(\b k)} = N^{- (|\del V| - n_\del)} D$ and we find
\bel{
  S_V^e[\trm{topo}] = (|\del V| - n_\del) \log N.
}
This result (in particular, the \emph{topological entanglement entropy} piece $-n_\del \log N$) was famously found in the early papers \cite{Hamma:2005zz, Kitaev:2005dm, Levin:2006zz}. It was also explicitly calculated in \cite{Casini:2013rba, Donnelly:2011hn} for the cubic lattice with $d = 2$, $N = 2$, and it is also precisely the result found using completely different tools by \cite{Yao:2010} while studying the topological state of the $\Z_2$ gauge theory on a honeycomb lattice. We here see that the  topological (i.e.~loop group-invariant) state in any $\Z_N$ gauge theory has the above entanglement entropy, as long as all superselection sectors have the same dimensionality (there are no defects on the lattice, etc). This result shows that in a topological state there always exists an ``area law'' piece of the entropy, which is lattice-dependent and will diverge in the continuum limit, and a finite piece which is sensitive only to the global data (number of disconnected components).

\section{Entanglement with magnetic boundary conditions}\label{sec mbc}

\subsection{Buffer zones}

So far, our discussion was largely parallel to \cite{Casini:2013rba}. We introduce magnetic boundary conditions rather differently. The key idea is to start from the setup with purely electric boundary conditions (where links are split into two disjoint sets $V$ and $\bar V$) and to modify the algebra $\A_V^e$ step-by-step by introducing magnetic boundary generators while keeping the number of superselection sectors constant. This way we will generate a family of algebras $\A_V^\alpha \subseteq \A_V^e$, and for each of them we will construct a reduced density matrix $\rho_V^\alpha$ such that $\Tr_{\H_V}(\rho_V^\alpha \O) = \avg \O$ for all $\O \in \A_V^{\alpha}$. We will always be ultimately interested in calculating $S_V^\alpha$, the von Neumann entropy of $\rho_V^\alpha$.

\begin{figure}
\begin{center}
\begin{tikzpicture}[scale = 2]
  \foreach \x in {-1, -0.5,..., 0.5} \draw (\x, 0) node[lightgray] {$\bullet$};
  \draw[step = 0.5, dotted] (-1.4, 0) grid (0.9, 1.4);
  \draw[step = 0.5] (-1.4, -0.9) grid (0.9, 0);

  \draw (3, 0) node[lightgray] {$\bullet$};
  \draw (4.5, 0) node[lightgray] {$\bullet$};
  \draw[step = 0.5, dotted] (2.6, 0) grid (4.9, 1.4);
  \draw[step = 0.5] (2.6, -0.9) grid (4.9, 0);

  \filldraw[fill=green!20, draw=green!50!black, thick]  (3.5, 0) rectangle +(0.5, 0.5);

  \draw[thick, blue, decorate, decoration = {snake, amplitude = .4mm, segment length = 2mm, post length = 0.1mm}] (-0.5, 0) -- (-0.5, 0.5);

  \draw[thick, teal, decorate, decoration = {snake, amplitude = .4mm, segment length = 2mm, post length = 0.1mm}] (0, 0) -- (0, 0.5);

  \draw[red] (-0.5, 0) node[anchor = north east] {$E_i$};
  \draw[green] (0, 0) node[anchor = north west] {$E_j$};
  \draw[blue] (-0.5, 0.5) node[anchor = north east] {$\bar E_i$};
  \draw[teal] (0, 0.5) node[anchor = north west] {$\bar E_j$};

  \draw[very thick, green] (0.5, -0.015) -- (-0.5, -0.015);
  \draw[very thick, green] (0, 0) -- (0, -0.5);
  \draw[very thick, red] (0, 0.01) -- (-1, 0.01);
  \draw[very thick, red] (-0.5, 0.01) -- (-0.5, -0.5);

  \draw (-0.25, 1.2) node[fill = white] {\textsc{before}};

  \draw [very thick, red] (3.5, 0) -- (3, 0);
  \draw [very thick, red] (4, 0) -- (4.5, 0);
  \draw [very thick, red] (3.5, 0) -- (3.5, -0.5);
  \draw [very thick, red] (4, 0) -- (4, -0.5);

  \draw[thick, blue, decorate, decoration = {snake, amplitude = .4mm, segment length = 2mm, post length = 0.1mm}] (3.5, 0.5) -- (3, 0.5);
  \draw[thick, blue, decorate, decoration = {snake, amplitude = .4mm, segment length = 2mm, post length = 0.1mm}] (3.5, 0.5) -- (3.5, 1);
  \draw[thick, blue, decorate, decoration = {snake, amplitude = .4mm, segment length = 2mm, post length = 0.1mm}] (4, 0.5) -- (4.5, 0.5);
  \draw[thick, blue, decorate, decoration = {snake, amplitude = .4mm, segment length = 2mm, post length = 0.1mm}] (4, 0.5) -- (4, 1);

  \draw[red] (3.75, -0.5) node[anchor = south] {$E_{ij}$};
  \draw[blue] (3.75, 1) node[anchor = north] {$\bar E_{ij}$};
  \draw[olive] (3.75, 0.25) node {$W_{ij}$};

  \draw (3.75, 1.2) node[fill = white] {\textsc{after}};
\end{tikzpicture}
\end{center}
\caption{\small (\textsc{color online}) A $d = 2$ example of one insertion of a magnetic boundary operator at link $(i, j)$ into a partition with purely electric boundary conditions. \emph{To the left:} as before, solid black lines are links in $V$,  dashed black lines are links in $\bar V$, and gray circles are elements of $\del V$. The product of electric generators on the red solid links gives the boundary electric operator $E_i$, and corresponding products on green solid links gives $E_j$. The inverses of these operators are respectively equal to $\bar E_i$, the product of electric generators on blue wavy links, and to $\bar E_j$, the product of electric generators on teal  wavy links. \emph{To the right:} Solid black lines are links in $V'$, dashed black lines are links in $\bar V'$, the shaded green plaquette is the newly inserted buffer zone, which (together with the gray circles) forms the new boundary $\del V'$. Electric generators on the solid red links multiply to give $E_{ij} = E_i E_j$, the new boundary electric operator that measures the electric flux flowing into the buffer zone. Its inverse is equal to $\bar E_{ij} = \bar E_i \bar E_j$, which is in turn given by the product of electric generators on the blue wavy links. The magnetic flux through the buffer zone placed at link $(i, j)$ is measured by $W_{ij}$, the magnetic generator around the green shaded buffer plaquette.}
\label{fig bdry magn op}
\end{figure}
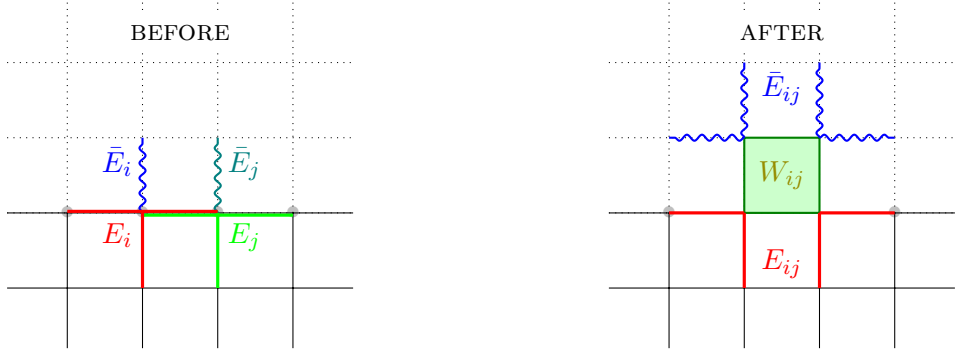

Let us start from the bipartition $(V, \bar V)$ and focus on two vertices $i, j \in \del V$ connected by exactly one link. Focus on one plaquette $p$ that contains $i$ and $j$ and has some links in $V$ and some links in $\bar V$. (If there are multiple such plaquettes, pick one.) Let us now define two new algebras, $\A_V^{(ij)}$ and $\A_{\bar V}^{(ij)}$, as the algebras generated by all the same generators that generated $\A_V$ and $\A_{\bar V}$ \emph{except} for the electric operators on links in $p$. This way we also define two new regions, $V'$ and $\bar V'$, as the sets of links in $V$ and $\bar V$ from which no electric generators were removed.  This prescription works in any dimension, as does our entire discussion up to now; Fig.~\ref{fig bdry magn op} shows a particularly clear example in $d = 2$.

The newly introduced sets of links form a tripartition $(V', \bar V', \del V')$ of the lattice. The third set in this tripartition is what we will call a \textbf{buffer zone}. In this case $\del V' = p$; by repeating the above procedure for other links in $\del V$ we put more plaquettes into $\del V'$. The magnetic generators on plaquettes in the buffer zone will be called \textbf{boundary magnetic generators}.

The new algebras evidently satisfy
\bel{
  \A_{V}^{(ij)} \subset \A_V,\quad \A_{\bar V}^{(ij)} \subset \A_{\bar V}.
}
The joint center $\mathcal Z_V^{(ij)}$ of algebras $\A_{V}^{(ij)}$ and $\A_{\bar V}^{(ij)}$ is generated by operators associated to elements of $\del V$ and $\del V'$, i.e.~by boundary electric generators on $V'$ and $\bar V'$ (respectively) \emph{and} by the magnetic generator on the remaining plaquette. This center has the same number of generators as the electric one. The algebra generated by all the operators in $\A_{V}^{(ij)} \cup \A_{\bar V}^{(ij)}$ differs from $\A$ by the absence of electric operators in the buffer zone.

In terms of boundary generators, the above procedure has taken the boundary electric generators $E_i$ and $E_j$ associated to two adjacent boundary vertices $i$ and $j$ and replaced them by a magnetic generator $W_{ij}$ on the plaquette containing both $i$ and $j$ and by an electric operator $E_{ij} = E_i E_j$ ensuring the overall conservation of electric flux through the entire buffer zone. The superselection sectors are now labeled by the eigenvalues of $W_{ij}$, $E_{ij}$, and all other $E_k$'s, i.e.~by (1) the magnetic flux flowing through the buffer zone, (2) the electric flux through all individual points of the boundary not touching a buffer zone, and (3) the electric flux through the buffer zone as a whole.

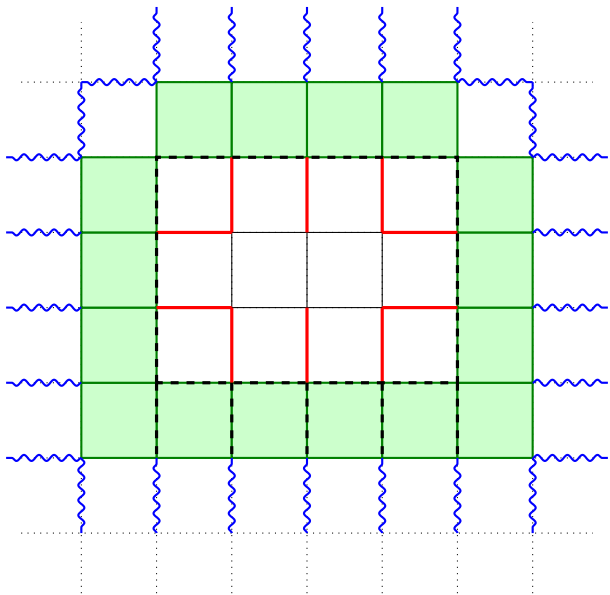
\begin{figure}[t]
\begin{center}
\begin{tikzpicture}[scale = 2]
  \draw[step = 0.5, dotted] (-1.9, -1.9) grid (1.9, 1.9);
  \draw[step = 0.5] (-1, -0.5) grid (1, 1);

  \foreach \x in {-1, -0.5,..., 0.5} \filldraw [fill=green!20, draw=green!50!black, thick] (\x, -1) rectangle +(0.5, 0.5);
  \foreach \x in {-1, -0.5,..., 0.5} \filldraw [fill=green!20, draw=green!50!black, thick] (\x, 1) rectangle +(0.5, 0.5);
  \foreach \x in {-1, -0.5,..., 0.5} \filldraw [fill=green!20, draw=green!50!black, thick] (-1.5, \x) rectangle +(0.5, 0.5);
  \foreach \x in {-1, -0.5,..., 0.5} \filldraw [fill=green!20, draw=green!50!black, thick] (1, \x) rectangle +(0.5, 0.5);

  \foreach \x in {-0.5, 0, 0.5} \draw [very thick, red] (\x, 1) -- (\x, 0.5);
  \foreach \x in {-0.5, 0, 0.5} \draw [very thick, red] (\x, -0.5) -- (\x, 0);
  \foreach \x in {0, 0.5} \draw [very thick, red] (1,\x) -- (0.5, \x);
  \foreach \x in {0, 0.5} \draw [very thick, red] (-1, \x) -- (-0.5, \x);

  \foreach \x in {-1, -0.5, 0, 0.5, 1} \draw [thick, blue, decorate, decoration = {snake, amplitude = .4mm, segment length = 2mm, post length = 0.1mm}] (\x, 1.5) -- (\x, 2);
  \foreach \x in {-1, -0.5, 0, 0.5, 1} \draw [thick, blue, decorate, decoration = {snake, amplitude = .4mm, segment length = 2mm, post length = 0.1mm}] (\x, -1.5) -- (\x, -1);
  \foreach \x in {-1, -0.5, 0, 0.5, 1} \draw [thick, blue, decorate, decoration = {snake, amplitude = .4mm, segment length = 2mm, post length = 0.1mm}] (1.5, \x) -- (2, \x);
  \foreach \x in {-1, -0.5, 0, 0.5, 1} \draw [thick, blue, decorate, decoration = {snake, amplitude = .4mm, segment length = 2mm, post length = 0.1mm}] (-1.5, \x) -- (-2, \x);
  \draw [thick, blue, decorate, decoration = {snake, amplitude = .4mm, segment length = 2mm, post length = 0.1mm}] (-1.5, -1) -- (-1.5, -1.5);
  \draw [thick, blue, decorate, decoration = {snake, amplitude = .4mm, segment length = 2mm, post length = 0.1mm}] (-1.5, 1) -- (-1.5, 1.5);
  \draw [thick, blue, decorate, decoration = {snake, amplitude = .4mm, segment length = 2mm, post length = 0.1mm}] (-1, 1.5) -- (-1.5, 1.5);
  \draw [thick, blue, decorate, decoration = {snake, amplitude = .4mm, segment length = 2mm, post length = 0.1mm}] (1.5, -1) -- (1.5, -1.5);
  \draw [thick, blue, decorate, decoration = {snake, amplitude = .4mm, segment length = 2mm, post length = 0.1mm}] (1.5, 1) -- (1.5, 1.5);
  \draw [thick, blue, decorate, decoration = {snake, amplitude = .4mm, segment length = 2mm, post length = 0.1mm}] (1, 1.5) -- (1.5, 1.5);

  \draw [dashed, very thick] (-1, -0.5) rectangle (1, 1);
  \foreach \x in {-1, -0.5, 0, 0.5, 1} \draw [dashed, very thick] (\x, -0.5) -- (\x, -1);
\end{tikzpicture}
\end{center}
\caption{\small (\textsc{color online}) A $d = 2$ example of a partition implementing purely magnetic boundary conditions. The ``outer links'' of $V$ in original, purely electric boundary conditions --- the ones that corresponded to the partition $(V, \bar V)$ on Fig.~\ref{fig el bdry ops} --- are shown by a black dashed line, and as before black solid lines correspond to links in $V'$ while black dotted lines correspond to links in $\bar V'$. The boundary $\del V'$ now consists solely of 16 buffer plaquettes shaded in green. The magnetic generators around these plaquettes generate the magnetic center. Both the product of electric generators on links leading into the buffer zone (solid red lines) and out of it (wavy blue lines) are identically unity, due to the global Gauss law constraint. At each prior step that lead from the purely electric to the purely magnetic boundary conditions, the number of generators of the center was kept constant.}
\label{fig magn bdry ops}
\end{figure}

This procedure can be iterated to obtain other algebras $\A_V^\alpha$. In particular, the \textbf{magnetic center} $\mathcal Z_{V}^m$ is the algebra generated solely by boundary magnetic generators. The relevant partitioning is achieved by starting from the usual $(V, \bar V)$ bipartition, removing plaquettes bordering links connecting adjacent vertices in $\del V$, and replacing them with magnetic generators until there is not a single vertex from which both a $V'$ link and a $\bar V'$ emanate. This way one obtains a single buffer zone $\del V'$ that completely insulates $V'$ and $\bar V'$ (Fig.~\ref{fig magn bdry ops}). The starting electric algebras $\A_V^e$ and $\A_{\bar V}^e$ are now reduced to the \textbf{magnetic algebras} $\A_V^m$ and $\A_{\bar V}^m$ that differ from their predecessors by the exclusion of all operators generated by electric links on $\del V$.

The electric eigenstates of the previous subsection will not be eigenstates of the boundary magnetic generators. The good quantum numbers are now the $|\del V'| = |\del V|$ magnetic fluxes $w_p$ through each plaquette in the buffer zone. The sum over superselection sectors is now the sum over all possible magnetic field configurations $\b w = \{w_p\}$ with $p \in \del V'$; these are the ``magnetic boundary conditions.'' The operators measuring the \emph{total} amount of electric flux passing into the buffer zone from $V'$ and $\bar V'$ are also in the magnetic center, but the Gauss law ensures that these operators must be equal to unity on all physical states, so they do not generate additional degrees of freedom. However, the product of all boundary magnetic generators is generated by operators solely in $\A_V^m \cup \A_{\bar V}^m$, which means that the center has only $|\del V| - n_\del$ independent elements, just like in the electric case.

The prescription given above is valid in any dimension, but is most easily visualized in $d = 2$. There exist additional minor subtleties in $d > 2$. Take, for instance, the cubic lattice in $d = 3$. Pick a link connecting two elements of $\del V$. According to our rule, the entire plaquette that contains this link and is perpendicular (and external) to $V$ can be replaced by a buffer zone. A fully insulating buffer zone is achieved when exactly two links on each plaquette in $\del V$ are replaced by buffer zones. In this case there will be the same amount of center generators as in the electric case. There exist many ways to choose this partitioning, so there may seem to be an ambiguity here. This is not so, as any two maximal partitionings (choices of fully insulating sets of plaquettes) can be related to each other by using operators that live on the boundary $\del V$ (or on the ``other boundary,'' which consists of vertices that are all in $V'$).

\subsection{A simple example}

So far we have merely introduced a different way of labeling boundary states. We are now ready to compute the entanglement entropy between the sets of links $V$ and $\bar V$. This computation is essentially different from the previous one, as the gauge-invariant states on buffer plaquettes can be entangled themselves, and in general the density matrix cannot be reduced to a block-diagonal form. Moreover, tracing out $\bar V$ links that lie in $\del V'$ merits more comments on its own. We focus on the latter first, and we study a very simple example.

\begin{figure}
\begin{center}
\begin{tikzpicture}[scale = 2]
  \fill[fill = green!20] (0, 0) rectangle (1, 1);
  \draw[dotted, thick] (0, 0) -- (0, 1) -- (1, 1) -- (1, 0);
  \draw[thick] (0, 0) -- (1, 0);
  \draw (0.5, 0) node[anchor = north] {$n_{\trm D}$};
  \draw (0.5, 1) node[anchor = south] {$n_{\trm U}$};
  \draw (0, 0.5) node[anchor = east] {$n_{\trm L}$};
  \draw (1, 0.5) node[anchor = west] {$n_{\trm R}$};
\end{tikzpicture}
\end{center}
\caption{\small The simplest lattice which supports magnetic boundary conditions. The solid line is the single link in $V$, and the dotted lines are the three links in $\bar V$. The entire plaquette makes up the buffer zone $\del V'$ (shaded region). The set of $V$ links not in the buffer zone ($V'$) and the set of $\bar V$ links not in the buffer zone ($\bar V'$) are both empty. The eigenvalues of the position operators $U_\ell$ on the four links $\ell \in \{\trm U, \trm D, \trm L, \trm R\}$ are $n_\ell$. }
\label{fig ex plaquette}
\end{figure}
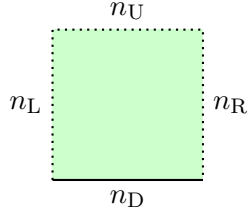

Consider a square $d = 2$ lattice with $N_P = 1$, i.e.~a lattice with four links and only one plaquette (Fig.~\ref{fig ex plaquette}). Label the links L, R, D, and U, and choose the partition $V = \{\trm D\}$, $\bar V = \{\trm R, \trm L, \trm U\}$. In the purely electric basis, the physical states are elements of an $N$-dimensional Hilbert space spanned by basis vectors $\qvec k$ of states with electric flux $k$ running along this plaquette. There are two boundary electric operators and they have eigenvalues $k$ and $-k$ in the electric basis. The topological state is $\qvec{\trm{topo}} = \frac1{\sqrt N} \sum_{k = 0}^{N - 1} \qvec{k}$ and in it the entanglement entropy between $V$ and $\bar V$ is given by \eqref{SV topo}, yielding $S_V^e = \log N$.

To get the magnetic boundary conditions, we turn this plaquette into a buffer zone. The states of objects living in the buffer zone are expressed in the magnetic basis, which is formed by the eigenstates $\qvec w$ of the magnetic generator $W$ around the buffer plaquette. In this basis, the topological state is $\qvec{\trm{topo}} = \qvec{0}$. Thus we are in the $w = 0$ superselection sector. Since $V' = \bar V' = \varnothing$ (all links belong to the plaquette in $\del V'$), the entropy associated to integrating out degrees of freedom in $V'$ is trivially $S_{V'} = 0$ in any state, including $\qvec{\trm{topo}}$. It is no surprise that $S_V = \log N$ is not reproduced. To measure this entanglement, we must integrate out the remaining $\bar V$ degrees of freedom by resolving the gauge-invariant state $\qvec 0$ into an entangled pair of gauge-variant states living on $V$ and $\bar V$.\footnote{This will not violate gauge invariance because the relevant density matrix will always make sure to superimpose gauge-variant states in a gauge-invariant manner. This is equivalent to the following setup. Consider two particles in the spin-$\frac12$ representation of $SU(2)$. We may start from their singlet state and trace over the Hilbert space of one particle. The spin eigenstates of the remaining particle are certainly not $SU(2)$ singlets, but the reduced density matrix we get must be such that only $SU(2)$-invariant expectations will be nonzero. In particular, the von Neumann entropy of this matrix gives the entanglement entropy between the two particles.} Harking back to \eqref{pos mom ops}, let $\qvec {g^{n_\ell}}_\ell$ be a position eigenstate on link $\ell$. The states of the buffer zone in $V$ (i.e.~the link D) are spanned by position eigenvectors $\qvec {g^{n\_D}}_V$, and the states on the buffer links in $\bar V$ are spanned by the tensor product of the position eigenvectors on the remaining three links, i.e.~by $\qvec {g^{n\_L},\,g^{n\_R},\,g^{n\_U}}_{\bar V}$. For the purposes of calculating entanglement entropy, it is enough to consider eigenstates $\qvec{g^{n_{\bar V}}}_{\bar V}$ of $U\_R U\_U\+ U_L\+ $ labeled by a single number $n_{\bar V} = n\_R - n\_U - n\_L$; then any state $\qvec w$ can be written as the singlet projection (sum over all gauge orbits) $\qvec w= \frac1{\sqrt N} \sum_{n = 0}^{N - 1} \qvec{g^n}_V \qvec {g^{w - n}}_{\bar V}$. The topological state $\qvec 0$ thus entangles the $V$ and $\bar V$ states with density matrix $\rho^{(w = 0)} = \frac1N \sum_{n, m} \qvec{g^n}_V \qvec {g^{-n}}_{\bar V} \qvecconj{g^m}_V \qvecconj{g^{-m}}_{\bar V}$. Tracing out the $\bar V$ states leaves the reduced density matrix $\rho_V^{m} = \frac1N \sum_n \qvec{g^n}_V \qvecconj{g^n}_V$ whose von Neumann entropy is precisely $S_V^m = \log N$.

As an additional check, let us consider the vacuum state, i.e.~the state without any electric flux lines. In the magnetic basis, it is given by $\qvec\Omega = \frac1{\sqrt N} \sum_{w = 0}^{N - 1} \qvec w$. In terms of the $V$ and $\bar V$ states, the density matrix is $\rho = \frac1{N^2} \sum_{n, m, w, v} \qvec {g^n}_V \qvec{g^{w - n}}_{\bar V} \qvecconj {g^m}_V \qvecconj{g^{v - m}}_{\bar V}$. Integrating out the $\bar V$ states, the reduced density matrix is $\rho_V^m = \frac1N \sum_{n, m} \qvec{g^n}_V \qvecconj{g^m}_V$, so this is a pure state with von Neumann entropy $S_V^m = 0$, as expected.

Heuristically, the difference between $S_{V'}$ and $S_V^m$ shows that since the magnetic basis uses eigenvalues of non-local operators, some very short-range entanglement can be lost due to this non-locality. If we insisted on working with degrees of freedom living on plaquettes and not on links, asking to quantify the entanglement between sets of links would have been as ambiguous or ill-defined as asking to quantify the entanglement between sets of vertices in a gauge theory.

A closer comparative study of the electric and magnetic computations above shows that the entire difference between the two lies in using a different basis for the Hilbert space. Since bases can be chosen independently for links in $V$ and $\bar V$, it is in a way natural that we have obtained the same answers. We will put this idea on firmer footing in Sec.~\ref{sec embc}. At this point, however, we remark that there are indications that this basis change will impact the end result in non-Abelian lattice gauge theories, and that this basis-dependence appears related to the electric-magnetic duality properties of the theory in question \cite{Jian:2014}.  We leave these questions to future research.

\subsection{The general case}

Having understood this almost trivial example, it is straightforward to generalize to the case with an arbitrary lattice, an arbitrary partition $(V, \bar V)$, and an arbitrary density matrix. Let us focus on purely magnetic boundary conditions; mixed boundary conditions are treated analogously. As in the example above, our goal is to find the entanglement entropy $S_V^m$ associated to the original partition. To do this, we must resolve the gauge-invariant Wilson loop eigenstates in the buffer zone $\del V'$ into eigenstates of position operators $U_\ell$ lying purely in $V$ and $\bar V$. As this calculation is not as neat as the electric one, we proceed in a series of elementary steps.

We work in the magnetic basis on $\del V'$, wherein the magnetic eigenstate with $\b w$ flux is denoted $\qvec{\b w}_{\del V'}$.  A density matrix $\rho$ on $\H$ is written as
\bel{\label{magn basis rho}
  \rho = \sum_{\substack{\b w,\, \b w' \in \H_{\del V'}\\ U,\, U' \in \H_{V'}\\ \bar U,\, \bar U'\in \H_{\bar V'}}} \rho_{\b w\, U\, \bar U,\ \b w'\, U'\, \bar U'} \qvec{\b w}_{\del V'} \qvecconj{\b w'}_{\del V'}\ \qvec U_{V'} \qvecconj {U'}_{V'} \ \qvec {\bar U}_{\bar V'} \qvecconj {\bar U'}_{\bar V'}.
}
Here, $\H_{\bar V'}$ is the space of states on links in $\bar V'$. In the electric basis for $\H_{\bar V'}$, the above sum runs over all states with zero total electric flux entering the buffer zone. These states are the eigenstates of the product of all electric operators on wavy blue lines on Fig.~\ref{fig magn bdry ops} with eigenvalue one. Similarly, the $U$'s in \eqref{magn basis rho} can be thought of as all electric eigenstates on links in $V'$ with zero net flux flowing into the buffer zone, and the $\b w$'s are the different magnetic fluxes running through boundary plaquettes.

Integrating out all degrees of freedom from $\bar V$ that are not in the buffer zone (that is, all degrees of freedom on $\bar V'$, or all degrees of freedom not in $V' \cup \del V'$), we find the reduced density operator
\bel{\label{reduced rho}
  \rho_{V' \cup \del V'} \equiv \Tr_{\H_{\bar V'}} \rho.}
We must also trace out the degrees of freedom living on $\bar V$ links that are in the buffer zone $\del V'$. As in the simple example above, this can be done plaquette-by-plaquette. Take a magnetic eigenstate $\qvec{w_p}_p$ on a buffer plaquette $p$, and let $p_V$ and $p_{\bar V}$ be the links in $p$ that are in $V$ and $\bar V$, respectively, and not in any other buffer plaquette.  Let $\qvec{g^{n}}_{p_V}$ be the eigenstates of the product of position operators along $p_V$, and analogously for $\qvec{g^{n}}_{p_{\bar V}}$. The magnetic eigenstate must be given by a gauge-invariant superposition of direct products of these two states, namely
\bel{\label{wp}
  \qvec{w_p}_p = \frac1{\sqrt N} \sum_{n_p = 0}^{N - 1} \qvec{g^{n_p}}_{p_V} \qvec{g^{w_p - n_p}}_{p_{\bar V}}.
}
By doing this, we have resolved the Hilbert space of states in $\del V'$ by adding $N$ degrees of freedom at each of $|\del V'| - n_\del$ plaquettes. The space of states on the subset $V$, $\H_V$, is spanned by the electric flux loops on $V'$ with zero outbound electric flux (i.e.~the basis vectors of $\H_{V'}$) and by the position eigenstates of links on the $p_V$'s. Note that $\H_V$ is the same Hilbert space as defined in \eqref{HV}; we just use a different basis for the degrees of freedom on $\del V$.

With $\qvec{w_p}_p$ given by \eqref{wp}, we can substitute $\qvec{\b w}_{\del V'} = \bigotimes_{p \in \del V'} \qvec{w_p}_p$ into \eqref{magn basis rho} and trace over all the $p_{\bar V}$ degrees of freedom to find the reduced density matrix in the magnetic basis,
\bel{\label{magn rho v}
  \rho_V^m = \frac1{N^{|\del V'| - n_\del}} \sum_{\substack{\b n,\, \b w,\, \b w'\\ U,\, U'}} \left(\rho_{V' \cup \del V'}\right)_{\b w\, U,\ \b w'\, U'} \qvec{\b n} \qvecconj{\b w' - \b w + \b n}\ \qvec U_{V'} \qvecconj {U'}_{V'},
}
where $\left(\rho_{V' \cup \del V'}\right)_{\b w\, U,\ \b w'\, U'}$ is the matrix element of the reduced matrix defined in \eqref{reduced rho}, and where $\b n = \{n_p\}$ labels the basis of the states on the $p_V$'s,
\bel{
  \qvec{\b n} = \bigotimes_{p \in \del V'} \qvec{g^{n_p}}_{p_V}.
}
The density matrix does not factorize into block-diagonal form as it did in the electric case. However, we can define
\bel{
  \rho_V^m \equiv \sum_{\b w} p_{\b w} \left( \rho_V^{(\b w)} \otimes \b J_{\b w}\right).
}
Here we have defined the unit matrix offset horizontally by $\b w$,
\bel{
  \b J_{\b w} \equiv \sum_{\b n} \qvec{\b n} \qvecconj{\b n + \b w},
}
and
\bel{
  p_{\b w} \left(\rho_V^{(\b w)}\right)_{U\, U'} \equiv \frac1{N^{|\del V'| - n_\del}} \sum_{\b w'} \left(\rho_{V' \cup \del V'}\right)_{\b w',\, U,\, \b w' + \b w,\, U'}
}
with
\bel{
  p_\b w \equiv \frac1{N^{|\del V'| - n_\del}} \sum_{\b w',\, U} \left(\rho_{V' \cup \del V'}\right)_{\b w',\, U,\, \b w' + \b w,\, U}
}
ensuring that the $\rho_V^{(\b w)}$'s have unit trace. Note that $p_{\b 0} = 1/N^{|\del V'| - n_\del}$.

The entanglement entropy is, as usual, the von Neumann entropy
\bel{
  S_V^m = - \Tr_{\H_V} \left( \rho_V^m \log \rho_V^m \right).
}
Note that, in general, there is no simple expression for the entanglement entropy with magnetic boundary conditions, in stark contrast with the compact formula \eqref{SV} found in the electric case.

To illustrate this on a simple example, let us pick the topological state with $\b w = \b 0$ everywhere. Now $\b J_{\b 0}$ is a diagonal matrix and $p_{\b 0} = 1/N^{|\del V'| - n_\del}$, and we find
\bel{
  \rho_V^m[\trm{topo}] = \frac1{N^{|\del V'| - n_\del}} \1_{N^{|\del V'| - n_\del}} \otimes \left(\rho_{V'}\right),
}
where $\left(\rho_{V'}\right)_{U\, U'} = \left(\rho_{V' \cup \del V'}\right)_{\b 0\, U,\, \b 0\, U'}$ is the density matrix of the pure state with $w = 0$ on all plaquettes in $V'$. Thus, in the magnetic basis and within the topological state, once all non-$V$ degrees of freedom are integrated out, we are left with randomly fluctuating, uniformly distributed variables on the boundary links $\del V$. The associated entanglement entropy is easily found to be
\bel{
  S_V^m[\trm{topo}] = (|\del V'| - n_\del) \log N,
}
leading to the same entanglement entropy (and in particular to the same universal term $-n_\del \log N$) found using electric boundary conditions \eqref{SV topo}. The same result holds for any pure state that is the eigenstate of the boundary magnetic operators.

\section{A unified approach to electric and magnetic boundary conditions} \label{sec embc}

The work in the previous two sections essentially follows an algorithm for computing entanglement entropy in a given $\Z_N$ lattice gauge theory:
\begin{enumerate}
  \item Pick a set of links $V$.
  \item Find the electric algebra $\A_V^e$ generated by \emph{all} gauge-invariant operators on $V$.
  \item If desired, pick an alternative algebra $\A_V^\alpha$ by excluding electric generators on links in $\del V$. Excluding all electric operators on $\del V$ gives the magnetic algebra $\A_V^m$.
  \item Identify the center $\mathcal Z_V^\alpha$ of the chosen algebra $\A_V^\alpha$. It is generated by remaining boundary electric operators and by magnetic operators on plaquettes containing links that used to house electric generators.
  \item In the basis that diagonalizes all generators of $\mathcal Z_V^\alpha$, construct the reduced density matrix $\rho^\alpha_V$.
  \item The von Neumann entropy of $\rho^\alpha_V$ is the entanglement entropy $S_V^\alpha$.
\end{enumerate}
In steps 1--3, corresponding actions are performed on the complement $\bar V$; $\mathcal Z_V^\alpha$ is found as the non-trivial joint center of $\A_V^\alpha$ and $\A_{\bar V}^\alpha$. For step 3, note that excluding any other boundary operators from $\A_V^m$ results in effectively removing an entire link from $V$, so these are \emph{all} algebras we may construct given $V$.

In the examples we have studied, the choice of algebra $\A_V^\alpha$ (and of the naturally corresponding center $\mathcal Z_V^\alpha$) did not influence the entanglement entropy. It merely made us adopt a different basis in which to express the good quantum numbers. Indeed, we have already glimpsed traces of the idea that the choice of algebra is related to just having to use a different basis for the same Hilbert space $\H_V$ defined in \eqref{HV}. We now formalize this notion.

Let us focus again on the superselection sectors present in the partition $(V, \bar V)$ with purely electric boundary conditions. We have shown that these sectors can be labeled by the outgoing electric fluxes $k_i$ at each site $i \in \del V$. In the electric basis of gauge-invariant states (spanned by eigenstates of products of electric generators along each plaquette, cf.~eq.~\eqref{H}), and given the electric fluxes on the buffer links wholly in $V$ and $\bar V$, the $k_i$'s are completely determined by the constant fluxes flowing along the plaquettes that are neither wholly in $V$ nor wholly in $\bar V$. These are nothing more than the buffer plaquettes.

Thus, both electric and magnetic boundary conditions (and all the mixed ones in between) can be realized by splitting the lattice into two completely disjoint sets of links, $V'$ and $\bar V'$, such that they are separated by a buffer zone as in Fig.~\ref{fig magn bdry ops}. The choice of which boundary conditions we are working with is now simply the choice of whether we wish to label superselection sectors by eigenvalues of the magnetic flux through the buffer plaquettes or by eigenvalues of the electric flux along the buffer plaquettes. (Recall that the magnetic flux operator on a plaquette $p$ is the magnetic generator $W_p$, while the electric flux operator on that plaquette is $\prod_{\ell \in p} L_\ell$; eigenstates of the two are Fourier transforms of each other.) In particular, this implies that the matrices $\rho_V^\alpha$ should all be related by unitary transformations that implement basis changes, and hence their von Neumann entropies should all be equal,
\bel{
  S_V^\alpha = S_V.
}
This means that even though we have defined $\rho_V^\alpha$ so that $\Tr_{\H_V} \left(\rho_V \O\right) = \avg \O$ just for $\O \in \A_V^\alpha$, any $\rho_V^\alpha$ will correctly calculate expectations for all operators in $\A_V^e$, the maximal algebra of operators on links in $V$. We have just shown that \emph{any} algebra that can be associated to $V$ gives the same entanglement entropy.

It should be noted that the original construction of the electric center given in \cite{Casini:2013rba} is much closer to the view we adopt in this section. In their construction, the electric center operators lived on links that were neither in $V'$ nor in $\bar V'$; these are the links denoted by blue wavy lines in Fig.~\ref{fig el bdry ops}. The set of these links is a subset of our buffer zone $\del V'$. Instead of labeling electric fluxes flowing along the buffer plaquettes, we could have indeed chosen to label electric fluxes on just a subset $\delta V$ of links between adjacent buffer plaquettes. We do not take this approach, but it should be kept in mind that different bookkeeping choices like this may be used. In particular, the calculations of \cite{Levin:2006zz} and \cite{Donnelly:2011hn} are instances of using buffer links --- not entire buffer plaquettes --- and splitting them into gauge-invariant superpositions of gauge-variant states.

Thus, as soon as we choose a bipartition $(V, \bar V)$, we may identify the completely insulating buffer zones (maximal $\del V'$'s), and then all that remains is the choice of preferred bookkeeping (basis labeling, etc) for the degrees in freedom in these zones. The conclusion is that
\begin{center}
\noindent\emph{Entanglement entropies in lattice gauge theories are naturally associated to buffer zones, i.e.~to sets of plaquettes that fully insulate one part of the lattice from the other. }
\end{center}
In other words, unlike in theories of matter where entanglement boundaries are viewed as co\-di\-men\-sion-one surfaces in the spatial plane, in gauge theories the entanglement boundaries should be viewed as codimension-zero but minimally thin shells. The choice of electric, magnetic, or mixed boundary conditions is a choice of labeling the superselection sectors or of choosing center operators, and the entanglement entropy is not affected.  For a given choice of boundary conditions, there always exists an unambiguous way to split the degrees of freedom in the buffer zone and integrate out the ones supported on links in $\bar V$. This procedure just happens to be very different in different bases, as our calculations above have shown. For calculational convenience, a basis should be chosen such that the full density matrix is as close to diagonal as possible.

\section{$U(1)$ gauge theory} \label{sec u1}

Having exhaustively treated the $\Z_N$ theory, we now briefly turn to the compact $U(1)$ theory. The link variables $U_\ell$ now take values on an $S^1$, and they are naturally described by an angle $\phi_\ell \in [0, 2\pi)$ satisfying $U_\ell = e^{i\phi_\ell}$.  Non-Abelian gauge theories are expressed similarly, by writing $U_\ell = e^{i \phi_\ell^a T^a}$ for Lie algebra generators $T^a$, but we will only focus on the Abelian theory, where all electric operators can be diagonalized simultaneously.

The electric generators now act as
\bel{
  L_\ell \qvec{e^{i\phi}}_\ell = i \pder{}{\phi} \qvec{e^{i\phi}}_\ell,
}
while magnetic generators $W^r_p$ act as shown earlier in \eqref{em gens}. The electric eigenstates are
\bel{
  \qvec{k}_\ell = \int_0^{2\pi} \frac{\d \phi}{2\pi} e^{ik\phi} \qvec{e^{i\phi}}_\ell,\quad k \in \Z,
}
and the situation is quite analogous to the $\Z_N$ theory (whose limit $N \rar \infty$ heuristically reproduces the $U(1)$ theory). In particular, the entire discussion of electric and magnetic boundary conditions, superselection sectors, and the entanglement entropy transfers wholesale with the single caveat that the Hilbert space on each link is infinite-dimensional because $k$ takes all integer values, and hence the number of superselection sectors is infinite from the outset. We thus conclude that the entanglement entropy of a region $V$ insulated by a buffer zone is still given by \eqref{SV},
\bel{
  S_V = - \sum_{\b k} p_{\b k} \log p_{\b k} + \sum_{\b k} p_{\b k} S_V^{(\b k)},
}
with $\b k \in \Z^{|\del V| - n_\del}$ being electric boundary conditions. Notice that in this case, the entanglement entropy of the topological state \eqref{SV topo} diverges logarithmically. This is a UV divergence associated to short distances on the target manifold ($S^1$) and it should be present in all gauge theories with continuous (Lie) groups. We believe this is the gauge theory analogue of the divergence found in \cite{Agon:2013iva}.

\section{Gauge theories with matter} \label{sec matter}

Adding a matter sector is also straightforward. Let us return to studying the $\Z_N$ theory, and let us consider fundamental matter degrees of freedom that live on sites and take values in $\Z_N$. At each site there is now a Hilbert space isomorphic to $\C^N$ with an operator algebra generated by the matter position and momentum operators that act as
\bel{
  \pi_i \qvec{g^n}_i = \qvec{g^{n + 1}}_i,\quad \varphi_i \qvec{g^n}_i = e^{2\pi i n/N} \qvec{g^n}_i.
}
Of course, position operators in other representations can be considered, but we focus on this one. The gauge transformation acts on matter states as $\qvec{g^n}_i \mapsto \qvec{g^{n + \Lambda_i}}_i$, and the gauge-invariant operators are the charge operator $\pi_i$ and the Wilson line operator $W_{ij} = \varphi_i\+ U_{ij} \varphi_j$. In the charge (i.e.~electric) basis, the charge operators are diagonalized on each site, and the eigenstates $\qvec{q}_i = \sum_n e^{-2\pi i nq/N}\qvec{g^n}_i $ satisfy
\bel{
  \pi_i \qvec q_i = e^{2\pi i q/N} \qvec q_i, \quad \varphi_i \qvec q_i = \qvec{q - 1}_i.
}
In the electric basis, the gauge-invariant states at each link satisfy $\sum_\mu L_{i, \mu} \pi_i = 1$; the Gauss law is the usual statement that the sum of electric fluxes entering each site must be equal to the charge on that site.

Let us immediately approach partitioning the lattice by introducing a fully insulating buffer zone separating regions $V$ and $\bar V$. This buffer zone now consists of links \emph{and} sites. Superselection sectors corresponding to electric boundary conditions are labeled by the eigenvalues of charge operators on sites and of electric flux operators on plaquettes in the buffer zone. For instance, on a square lattice in $d = 2$, each buffer plaquette is labeled by five eigenvalues, four charges at the square vertices and the electric flux along the entire perimeter; two of the five are shared with adjacent buffer plaquettes. Magnetic boundary conditions on the same plaquette entail diagonalizing the four Wilson line operators on the sides of the square and the product of electric flux operators on the buffer plaquette and the two adjacent non-buffer plaquettes, if the adjacent plaquettes are labeled by electric boundary conditions. The two extreme choices of boundary conditions thus entail labeling superselection sectors by all electric flux (glueball) and matter charge configurations in the buffer zone (purely electric boundary conditions) or by all Wilson lines in the buffer zone (purely magnetic boundary conditions). Of course, in this case the reduced density matrix will not be block-diagonal, even in the electric basis, but the same calculation as in Sec.~\ref{sec mbc} can be carried out.

Note that in the case of a pure matter theory, all we need to do is to remove the Gauss law constraint and the electric flux (glueball) operators from the electric center. The remaining labels of superselection sectors are then the charges on the vertices in the buffer zone. These can be factored into two sets and reabsorbed into traces over degrees of freedom, one set in $V$ and the other set in $\bar V$; therefore the above prescription for gauge field entanglement entropy correctly reproduces the usual entanglement entropy in theories with only matter.

\section{The continuum limit}\label{sec continuum}

Finally, we remark that the prescription given in these notes naturally generalizes to the continuum case. Whether the given lattice gauge theory has a continuum limit is a separate question, and it must be answered by carefully identifying any critical points in the parameter space of the given Hamiltonian. However, it is useful to think about the different boundary conditions in a continuum picture, so here we provide the guidelines to how this works.

As the lattice spacing is taken to zero, the buffer zone $\del V$ becomes infinitely thin and can be regarded as a codimension-one surface. In the case of electric boundary conditions, the sum over superselection sectors now becomes an integral over all possible electric fields $E_\perp$ perpendicular to $\del V$, or equivalently an integral over all configurations of glueballs confined to the interior of $\del V$. For purely magnetic boundary conditions, the sum over superselection sectors becomes an integral over all possible magnetic fields $\b B_\parallel$ tangential to $\del V$. For $d = 2$, $\b B_\parallel = B \b e_z$ is essentially a scalar quantity that measures the flux piercing the spatial plane along $\del V$. For $d = 2$, $\b B_\parallel$ is a vector and we may choose any of its components to label the superselection sectors. Mixed boundary conditions now mean that we choose different patches of $\del V$ and on each we impose either electric or magnetic conditions. The presence of matter forces us to include all charge configurations (for electric boundary conditions) or Wilson lines (for magnetic boundary conditions) on the entangling boundary. It is of note that the labels of superselection sectors precisely organize following the familiar boundary conditions of electrodynamics, where changes of fields $\Delta E_\perp$ and $\Delta \b B_\parallel$  between two environments are respectively related to the charge density and the surface current density on the boundary. The notion of $\Delta E_\perp$ labeling superselection sectors has already appeared in \cite{Agon:2013iva}.

What divergences can be associated to the continuum limit?  This question is generally out of the scope of these notes, but a glimpse into its answer is afforded by the formula \eqref{SV topo} for the entanglement entropy of the topological state. The non-extensive piece $-n_\del \log N$ will remain finite in the continuum limit. On the other hand, the area law term will take the form $\frac1\epsilon \trm{Vol}(\del V) \log N$ and will diverge as the lattice spacing $\epsilon$ is taken to zero. It is reassuring that the universal term (topological entanglement entropy) survives the continuum limit.

\section{Outlook}

We have provided a natural framework for thinking about entanglement in Abelian gauge theories. The ideas given here are a generalization and reformulation of many older results. The main take-away message of these notes is that entanglement entropy is naturally defined on a codimension-zero shell (the ``buffer zone'') between two spatial regions, and that the previous calculations of entanglement entropies in lattice gauge theories can be seen as computations with different basis choices for the degrees of freedom in the buffer zone. Explicit computation of the entropy is possible and very easy in certain cases of interest.

There exists a host of directions for further research. A natural extension of this work concerns non-Abelian theories and different representation choices of Wilson loops and matter fields. It would be fascinating to understand whether the entanglement entropy takes on a particularly simple form in the large $N$ limit. A further immediate association is the exploration of implications of our prescription to understanding the Ryu-Takayanagi formula; perhaps generalizations of the holographic entanglement entropy  of \cite{Hartnoll:2012ux} can be given a firmer interpretation, and perhaps the exact holographic mapping \cite{Qi:2013caa} can also be fruitfully applied to our lattice computations. Another alluring topic to which our results may be applied is the study of entanglement entropies in gravity theories, as discussed in e.g.~\cite{Solodukhin:2011gn, Callan:1994py, McGough:2013gka, Kabat:1995eq, Donnelly:2012st}.

It would also be interesting to study entanglement entropy from a purely lattice-based point of view, both analytically and numerically. For instance, one can ask whether entanglement entropy on its own can be used to detect different phases of quantum systems. It is also possible to repeat our general procedure on study lattices with non-trivial topology. We hope to return to these topics in the future.

\section*{Acknowledgments}

It is a pleasure to thank Shamik Banerjee, Chao-Ming Jian, Nima Lashkari, and Steve Shenker for valuable discussions. This work has been supported by the SITP and the NSF Graduate Research Fellowship.

\end{document}